# CoPreTHi: A Web tool which combines transmembrane protein segment prediction methods

V.J. Promponas, G.A. Palaios, C. Pasquier, J.S. Hamodrakas and S.J Hamodrakas

Department of Cell Biology and Biophysics, Faculty of Biology, University of Athens, Athens 157 01, Greece,
E-mail: vprobon@biology.db.uoa.gr, gpal@biology.db.uoa.gr, pasquier@cc.uoa.gr, ihamod@biology.db.uoa.gr, shamodr@cc.uoa.gr

**ABSTRACT:** CoPreTHi is a Java based web application, which combines the results of methods that predict the location of transmembane segments in protein sequences into a joint prediction histogram. Clearly, the joint prediction algorithm, produces superior quality results than individual prediction schemes. The program is available at http://o2.db.uoa.gr /CoPreTHi

**INTRODUCTION**

Membrane proteins are involved in a variety of important biological functions. Their biological activity depends primarily on their spatial conformation. The three-dimensional structure of a protein is usually determined by single-crystal X-ray crystallography. However, suitable single crystals are not easily produced for membrane proteins, therefore, experimental data at atomic or near-atomic resolution for membrane proteins are very few.

Since experimental findings indicate that all the necessary information for a protein to fold into its native structure is coded into its amino acid sequence (Anfinsen, 1973), several attempts have been made to predict the three-dimensional structure of a protein from sequence alone (Rost and Sander, 1994; Rost and Sander, 1996), but with limited success only. Also, recently, genomic sequences from different organisms have plenty of orphan ORF's, making more demanding the existence of accurate methods for the prediction of protein structure and function. In these case, even discriminating between globular and membrane proteins might be rewarding.

In the field of membrane proteins it is often very important to predict the location of transmembrane segments along the sequence, since these are the basic structural building blocks defining their topology.

Several successful prediction algorithms have been developed for membrane proteins, which sometimes not only predict transmembrane segments, but also topology and their secondary structure. For globular proteins, when predicting secondary structure, it has been claimed that combined prediction schemes provide a higher degree of accuracy than individual prediction methods (Schulz et al., 1974, Argos et al., 1976, Hamodrakas, 1988).

**METHOD**

In this report, we present CoPreTHi, a Web-based application that uses the results of some popular prediction methods freely accessed over the WWW:
- DAS: Cserzo et al., 1997,
- ISREC-SAPS: Brendel et al., 1992,
- PHD: Rost et al. 1995,
- SOSUI: Hirokawa et al., 1998,
- TmPred: Hofmann and Stoffel, 1993,
- TopPredII: von Heijne, 1992,

and a method developed by our group
- PRED-TMR: Pasquier et al., 1998

combining them into a joint prediction histogram, to predict the location of transmembrane segments in protein sequences. There is also the possibility for a user to submit the results of any other method in a specified format and to include them into the joint prediction histogram as well.

An amino acid residue predicted to be a part of a transmembrane domain by three or more methods is considered to be a residue inside a transmembrane region; thus, a combined prediction is obtained. Optionally, observed results can also be entered (e.g. the FT records of a SWISS-PROT entry: Bairoch and Apweiler, 1997). A reliability index Q (Chou and Fasman, 1978) and a correlation coefficient C (Matthews, 1975) are calculated in this case, to evaluate the accuracy of each prediction method separately, as well as of the joint prediction. All the results can be presented either in plain text (without any HTML tables, for Web Browsers without this capability) or in HTML mode, including a graphical representation.

**ARCHITECTURE**

CoPreTHi consists of three subprograms. The first, a Java program, creates the input form for the data (results of individual prediction methods). The second, a C program, performs all calculations and sends output to the third, which is also a Java program. This is responsible for the display of the results, draws a graphical representation and creates a table, containing details about individual prediction schemes, the joint prediction, and, optionally, their performance against the observed data (if available). Since input/output is performed with Java programs, the interface is user friendly.

In the main page of our server (at the URL: http://o2.db.uoa.gr/ CoPreTHi/Main.html) there are links to the methods mentioned above. Therefore, a user can run the individual methods separately and manually copy and paste each method's results into the input form of our tool as well as any observed results. The only extra information required to produce the joint prediction histogram for all the methods is the number of amino acids of the sequence, which should be entered manually in the 'seqlen' text area. The name of the sequence may optionally be entered in an appropriate text area, and it is displayed in the results page.

**RESULTS**

The predictions of the individual methods mentioned above, were tested, on a representative set of 155 sequences of transmembrane proteins with reliable topology, deposited in SWISSPROT (as described by Pasquier et al., 1998). For all individual methods and the joint prediction, a reliability index (Q) and a correlation coefficient (C) were calculated for each protein. Significant differences were found between the different prediction methods for some of these proteins, even for characteristic examples, such as Bacteriorhodopsin. Although this is true, the mean values of Q's and C's calculated by individual methods on the entire test set of 155 proteins, do not differ significantly, varying from approximately 86.4% to 89.3% (Q) and 0.71 to 0.78 (C), whereas the joint prediction gives 91.6% and 0.79 respectively. Clearly, the joint prediction algorithm, produces superior quality results on this set of 155 transmembrane proteins than individual prediction schemes.

## AVAILABILITY

CoPreTHi is freely available for use through the Internet at the URL: http://o2.db.uoa.gr/CoPreTHi. It can be executed over the World Wide Web on any Java compatible Web Browser. A list of the results obtained for the set of the 155 protein sequences used for our tests is also available at the URL: http://o2.db.uoa.gr/CoPreTHi/Results155/Results.html. Detailed help and useful comments are located at the URL: http://o2.db.uoa.gr/CoPreTHi/help.html

## REFERENCES

Anfinsen, C.B. (1973) Principles that govern the folding of protein chains. *Science*, **181**, 223-230

Argos, P., Schwartz, J. and Schwartz, J. (1976) An assessment of protein secondary structure prediction methods based on amino acid sequence. *Biochim. Biophys. Acta*, **439**, 261-273

Bairoch, A. and Apweiler, R. (1997) The SWISS-PROT protein sequence data bank and its new supplement TrEMBL. *Nucl. Acids Res.*, **25**, 31-36

Brendel, V., Bucher, P., Nourbakhsh, I., Blaisdell, B.E. and Karlin, S. (1992) Methods and algorithms for statistical analysis of protein sequences. *Proc. Natl. Acad. Sci. USA*, **89**, 2002-2006.

Chou, P.Y. and Fasman, G.D. (1978) Prediction of the secondary structure of proteins from their amino acid sequence. *Adv. Enzymol.*, **47**, 45-148

Cserzo, M., Wallin, E., Simon, I., von Heijne, G. and Elofsson, A. (1997) Prediction of transmembrane alpha-helices in procaryotic membrane proteins: the Dense Alignment Surface method. *Prot. Eng.*, **10(6)**, 673-676

Hamodrakas, S.J. (1988) A protein secondary structure prediction scheme for the IBM PC and compatibles. *CABIOS*, **4(4)**, 473-477

Hirokawa, T., Boon-Chieng, S. and Mitaku, S. (1998) SOSUI: Classification and Secondary Structure Prediction System for Membrane Proteins. *Bioinformatics (formerly CABIOS)*, **14(4)**, 378-379

Hofmann, K. and Stoffel, W. (1993) TMbase - A database of membrane spanning proteins segments. *Biol. Chem. Hoppe-Seyler*, **347**,166

Matthews, B.W. (1975) Comparison of the predicted and observed secondary structure of T4 phage lysozyme. *Biochim. Biophys. Acta*, **405**, 442-451

Mitaku, S., Boon-Chieng, S. and Hirokawa, T. (1998) A Theoretical Method to Distinguish between Membrane and Soluble Proteins by Physicochemical Approach (Submitted for publication)

Pasquier, C.M., Promponas, V.J., Palaios, G.A., Hamodrakas, J.S. and Hamodrakas, S.J. (1998) A novel method for predicting transmembrane segments in proteins based on a statistical analysis of the SwissProt database: the PRED-TMR algorithm. (Submitted for publication)

Rost, B., Casadio, R., Fariselli, P. and Sander, C. (1995) Prediction of helical transmembrane segments at 95% accuracy. *Prot. Sci.*,**4**, 521-533

Rost, B. and Sander, C. (1994) Structure prediction of proteins-Where are we now? *Curr. Opin. Biotechnol* ,**5**, 372-380

Rost, B. and Sander, C. (1996) Bridging the protein sequence-structure gap by structure predictions. *Annu. Rev. Biophys. Biomol. Struct.*, **25**, 113-136

Schulz, G.E., Barry, C.D., Friedman, J., Chou, P., Fasman, G.D., Finkelstein, A.V., Lim, V.I., Ptitsyn, O.B., Kabat, E.A., Wu, T.T., Levitt, M., Robson, B. and Nagano, K. (1974) Comparison of predicted and experimentally determined secondary structure of adenyl kinase, *Nature*, **250**, 140-142

von Heijne, G. (1992) Membrane Protein Structure Prediction, Hydrophobicity Analysis and the Positive-inside Rule. *J. Mol. Biol.*, **225**, 487-494